# Stability of a natural palygorskite after a cycle of adsorption-desorption of an emerging pollutant

D. Hernández[a], L. Quiñones[a], C. Charnay[b], M. Velázquez[c] y A. Rivera[a†]

[a] Institute of Materials Science and Technology (IMRE), University of Havana, Cuba.
[b] Institut Charles Gerhardt Montpellier (ICGM), Université Montpellier, France.
[c] Research Center for Mining Metallurgy Industries (CIPIMM), La Havana, Cuba.

[†] Corresponding author: aramis@imre.uh.cu

**Abstract**

In this paper, we evaluate the structural stability of a natural Cuban clay –an adsorbent of organic pollutants– after an adsorption/desorption process. The clay under study was palygorskite (Pal), and sulfamethoxazole (SMX) was the emerging contaminant. The materials were characterized by X-ray diffraction (XRD), attenuated total reflection (ATR) infrared spectroscopy and zeta potential (ZP) before and after the SMX adsorption/desorption processes. Based on the material integrity, the potentiality of Pal as pollutant adsorbent and its possible reuse in adsorption/desorption cycles was demonstrated.

**Introduction**

The use of different types of supports for organic species adsorption [1,2] in environmental applications is vital for decontamination. Special attention is given to naturals materials with adsorptive properties [3]. However, even when many authors work on the adsorption of organic pollutants by natural matrices, it is not common to investigate the physical and chemical properties of the matrices after being submitted to an adsorption-desorption cycle. The main motivation of this paper is to evaluate the structural stability of the natural clay palygorskite–an adsorbent of organic contaminants–after an adsorption–desorption process.

Natural Cuban clay type Palygorskite (Pal) is an excellent adsorbent of organic molecules [4]. It is a fibrous clay with structure 2:1 formed by tetrahedral silica sheets ($SiO_4^{4-}$) periodically inverted with respect to the tetrahedral bases leading to an interruption in the octahedral sheets ($AlO_3(OH)_3^{6-}$), and the formation of an open-channel structure [5-7]. Three water types are present in the structure (zeolitic, coordinated and structural water), as well as compensating cations in the tunnels [8,9]. From previous work [10] it is known that Pal acts like an effective adsorbent of emerging contaminants–chemical species of common used present in water resources which have the potential to cause adverse ecological and (or) human health effects [11]–such as sulfamethoxazole (SMX).

**Experimental**

In order to demonstrate the possibility to reuse the Pal support in adsorption-desorption cycles, the material resulting of the desorption process (labeled as Pal DE) was characterized by X-ray diffraction (XRD), attenuated total reflection (*ATR*) infrared spectroscopy and zeta potential (ZP). The X-ray diffraction (XRD) patterns for the samples Pal and Pal DE were conducted on a Philips Xpert diffractometer, using Cu-*Kα* radiation (λ = 1.54 Å) at room temperature for a range from 4 to 70°. ATR spectra were collected using a Perkin Elmer UATR Two FTIR Spectrometer in the range of 400-4000 cm$^{-1}$. The surface charge of clay particles was evaluated using a zeta potential analyzer (Malvern Nano Zetasizer instrument). For the analysis, 1 mg of the samples (Pal or Pal ED) was dispersed fully in 2 mL of a deionized water suspension at different pH (2 to 10) under ultrasonic stirring for 15 min. The measurements were made in triplicate and the averages were reported.

**Results and discussion**

As can be observed in figure 1, the main mineralogical phases present in the raw sample (Pal) and that resulting from desorption study (Pal DE) were palygorskite (P), montmorillonite (Mt), and quartz (Q) [12].

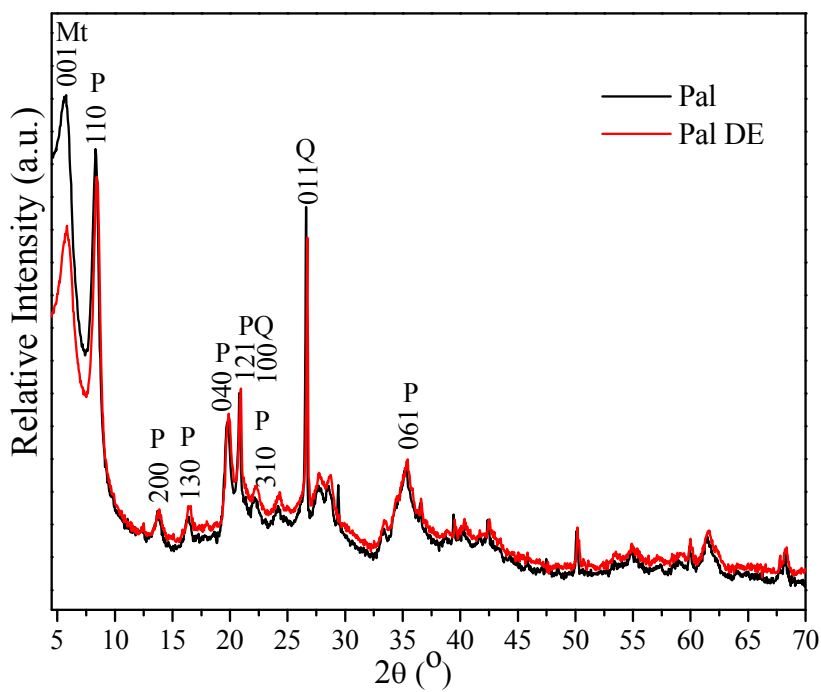

**Figure 1:** Diffraction patterns for Pal and Pal ED samples.

In the literature, the comparison between experimental and theoretical powder XRD patterns has allowed to note that most palygorskite samples are mixtures of monoclinic and or-thorhombic polymorphs [13,14]. Reflections corresponding to d-spacings between 4.0 and 4.5 Å, called "Chisholm zone", allow to determine if the palygorskite under study is mainly orthorhombic, monoclinic or either a mixture of both phases. However, in natural materials–as the Pal sample presented here–the presence in this zone of reflections associated to spurious phases can difficult this analysis.

No significant variations were observed in the interplanar distances values of the characteristic diffraction maximums of palygorskite between the Pal and Pal ED samples. Such result suggests no structural changes in the after an adsorption-desorption cycle, confirming its stability.

In addition, a decrease in the intensity of the 001 basal reflection of the montmorillonite phase was detected in the sample Pal DE. This might be due to possible dissolution of the Mt phase by acid pH effect during the SMX adsorption process.

ATR spectra of the organic molecule SMX (emerging contaminant from wastewater), the Pal-SMX composite, and the samples Pal and Pal DE are shown in figure 2. In the Pal-SMX spectrum, compared to the raw material (Pal), the appearance of signals associated with the presence of the organic molecule can be noted. However, for sample Pal ED these signals disappear as result of desorption process. As can be seen in the figure, very similar ATR spectra were obtained for the samples Pal and Pal ED (see figure 2): no variation in the characteristic bands of the clay were observed indicating the reversibility of the process as well as the possible reuse of the support material.

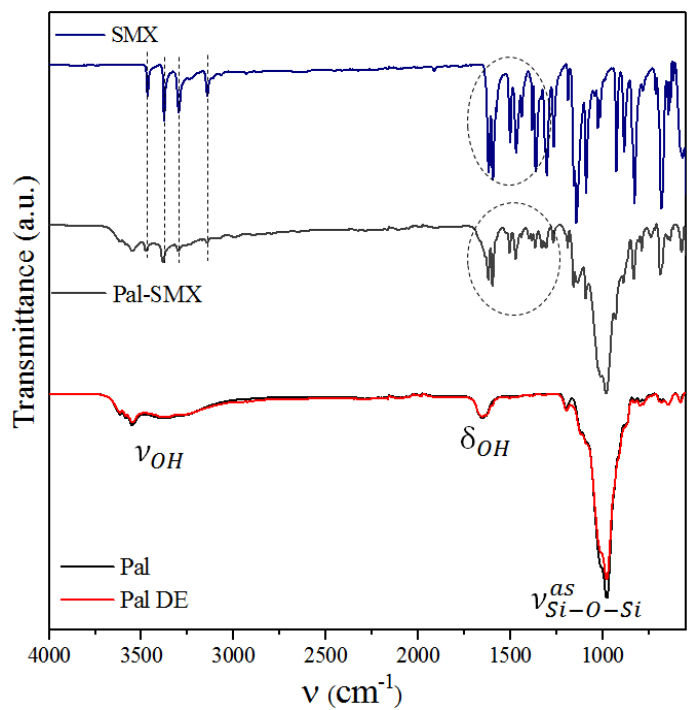

**Figure 2:** ATR spectra for samples of sulfamethoxazole (SMX), raw material (Pal), resulting composite of adsorption process (Pal-SMX) and material after the desorption process (Pal DE). In dotted lines the signals associated to the emerging pollutant.

The zeta potential results indicated that the surface charge of the samples PAL and Pal ED, before and after of the adsorption–desorption processes, was negative (≈−25 mV). A similar result has been reported in the literature for 2:1 clays [15,16]. In general, the quantitative and qualitative behavior obtained for both clay materials was basically the same.

**Conclusions**

The different analysis support the hypothesis that the Pal raw material can be reused as an efficient adsorbent of the emerging contaminant SMX, based on the total material integrity after the adsorption-desorption process.